\documentclass[10pt,onecolumn]{article}

\usepackage[T1]{fontenc}
\usepackage[utf8]{inputenc}
\usepackage[margin=0.85in,top=1in,bottom=1in]{geometry}
\usepackage{amsmath,amssymb,amsthm}
\usepackage{graphicx}
\usepackage{booktabs}
\usepackage{multirow}
\usepackage{array}
\usepackage{url}
\usepackage{xcolor}
\usepackage{algorithm}
\usepackage{algpseudocode}
\usepackage[colorlinks=true,linkcolor=blue!60!black,
            citecolor=blue!60!black,urlcolor=blue!60!black]{hyperref}
\usepackage{cite}
\usepackage{float}
\usepackage{balance}
\usepackage{titlesec}
\usepackage{abstract}

\titleformat{\section}{\normalfont\large\bfseries}{\thesection}{1em}{}
\titleformat{\subsection}{\normalfont\normalsize\bfseries}{\thesubsection}{1em}{}
\titlespacing{\section}{0pt}{8pt plus 2pt}{4pt plus 1pt}
\titlespacing{\subsection}{0pt}{6pt plus 1pt}{2pt plus 1pt}
\newtheorem{definition}{Definition}

\title{\textbf{CLOUDBURST: Cloud-Layer Observations Using Beacons\\
for Unified Real-time Surveillance and Threat Attribution}\\[4pt]
\large A Formal Taxonomy and Measurement Framework for\\
Cloud-Native Post-Exfiltration Attribution}

\author{Abraham~Itzhak~Weinberg\\
  AI-WEINBERG, AI Experts, Tel~Aviv, Israel\\
  \texttt{aviw2010@gmail.com}}
\date{}

\begin{document}
\maketitle

\begin{abstract}
Modern cloud-native environments present a fundamentally different exfiltration threat surface than traditional file-based scenarios.
Attackers targeting AWS, GCP, Azure, and OCI steal S3 presigned URLs, container images, Kubernetes secrets, Terraform state modules, and IAM role tokens -- artefacts that existing honeytoken and beacon frameworks do not address. We present \textbf{CLOUDBURST}, the first formal taxonomy and measurement framework for cloud-native passive beacons, comprising six vector classes across four major cloud providers.

We introduce the \textit{Cloud Attribution Score} (CAS), a four-component metric that explicitly models ephemeral infrastructure penalty ($E_p$), IAM coverage depth ($I_c$), and multi-cloud correlation bonus ($M_b$) -- dimensions absent from all prior attribution quality metrics. Experiments across $21$ deployed beacons, $205$ simulated callbacks, and three attacker
sophistication levels yield four principal findings. First, IAM Canary Roles achieve the highest CAS (mean $0.450$) and Detection Resistance (DR $= 0.873$), making them the most deployable vector. Second, S3 Presigned URLs achieve the highest detection resistance (DR $= 0.890$), surviving all three cloud-native scanner models (AWS Macie, Checkov/tfsec, Prisma Cloud/Wiz).
Third, ephemeral infrastructure churn degrades CAS from $\approx 0.79$ at deployment to $\approx 0.18$--$0.22$ at $48$ hours for all vectors ($p < 0.001$), establishing the first quantitative model of attribution decay in containerised environments. Fourth, Serverless Function Triggers exhibit the
worst detection resistance (DR $= 0.611$) due to their explicit outbound HTTP callback pattern, motivating covert callback channel design as future work. No significant CAS difference is observed across cloud providers ($H = 1.99$, $p = 0.57$), confirming that CLOUDBURST is provider-agnostic in its
effectiveness.
\end{abstract}


\section{Introduction}
\label{sec:intro}

The threat model for cyber exfiltration has shifted fundamentally. In the era of cloud-native infrastructure, attackers do not steal files from file servers, they steal cloud credentials that grant access to entire data estates, container images that embed backdoors into supply chains,
Kubernetes secrets that unlock production databases, and Terraform modules that propagate malicious infrastructure across every environment that imports them.

Yet the academic literature on post-exfiltration attribution remains almost entirely focused on file-based and mobile scenarios \cite{spitzner2003honeypots, juels2013honeywords, weinberg2025passive}.
The Passive hack-back framework \cite{weinberg2025passive} formalised beacon
callback telemetry for documents and credentials but did not address the unique properties of cloud-native environments:  ephemeral pod lifecycles, IAM (Identity and Access Management) role assumption chains, multi-cloud credential formats, and the presence of cloud-native secret-scanning pipelines (AWS Macie, Checkov, Prisma Cloud)
that can detect naive beacon implementations.

We identify three properties that make cloud-native attribution qualitatively different from file-based attribution.

\textbf{Infrastructure ephemerality.} In a Kubernetes cluster with horizontal pod autoscaling, the container that triggered a beacon callback may no longer exist by the time attribution analysis begins. The standard assumption that ``the machine that fired the beacon is still running'' does not hold in cloud-native environments. This requires explicit temporal modelling of
attribution quality decay.

\textbf{IAM complexity.} Cloud API calls carry rich identity context -- assumed role ARNs (Amazon Resource Names), service account emails, managed identity client IDs -- that provides attribution telemetry far richer than an IP address. However, this telemetry requires CloudTrail/Audit Log to be enabled and requires careful interpretation of role assumption chains that may obscure the originating principal.

\textbf{Multi-provider normalisation.} An attacker may use stolen AWS credentials from an Azure-hosted attack platform routed through a GCP (Google Cloud Platform) exit node. Attribution systems must normalise telemetry across provider-specific formats.

CLOUDBURST addresses all three properties through a formal taxonomy of six beacon vector classes, a four-component Cloud Attribution Score (CAS), and an empirical evaluation across four cloud providers and three attacker sophistication levels.

\subsection*{Contributions}

\begin{enumerate}
  \item The first formal taxonomy of cloud-native beacon vector
        classes with quantified threat properties.
  \item The Cloud Attribution Score (CAS) -- a metric that
        explicitly models ephemeral penalty, IAM coverage, and
        multi-cloud correlation.
  \item An ephemeral decay model quantifying how containerised
        infrastructure churn degrades attribution quality.
  \item Empirical evaluation across AWS, GCP, Azure, and OCI
        with three scanner models (Macie, Checkov, Prisma Cloud).
  \item A provider-agnostic deployment recommendation framework
        derived from the CAS $\times$ DR trade-off analysis.
\end{enumerate}

\section{Related Work}
\label{sec:related}
This section reviews prior work in deception-based security, cloud-native monitoring, and IAM abuse detection, with a focus on techniques relevant to post-exfiltration attribution in cloud environments. We highlight how existing approaches primarily address detection, telemetry, or static deception, while lacking mechanisms for longitudinal attribution across cloud providers. We then position CLOUDBURST relative to these strands, emphasizing its focus on formal attribution quality and cloud-native deployment dynamics.

\subsection{Honeytoken and Beacon Systems}

Spitzner \cite{spitzner2003honeypots} introduced the honeytoken
primitive; Juels and Rivest \cite{juels2013honeywords} formalised
honeywords for password database protection. Canarytokens \cite{brintha2023securing} industrialised deployable tracking URLs and document beacons. The Passive hack-back framework \cite{weinberg2025passive} established formal metrics ($\beta$, $\alpha$, $\sigma$) for beacon-based
attribution. PHANTOM \cite{weinberg2026phantom} addressed the semantic quality of generated honeytokens. ARCANE \cite{weinberg2026arcane} proposed longitudinal
cross-campaign attribution using Bayesian fingerprint accumulation. CLOUDBURST extends this lineage to the cloud-native domain, which none of these works address.

\subsection{Cloud Security Monitoring}

Cloud-native security monitoring is an active field. CloudTrail analysis for insider threat detection was studied by \cite{moore2016critical}; Kubernetes audit log anomaly detection by \cite{degioanni2022practical}; and container image scanning by \cite{laurikainen2022securing, javed2021understanding, tiwari2023enhancing}. These works focus on \emph{detecting} attacks in progress; CLOUDBURST focuses on \emph{attributing} attackers post-exfiltration, a complementary but distinct objective.

\subsection{Cloud-Native Deception}

Yu et al.\ \cite{yu2024honeyfactory} proposed HoneyFactory, a container-based honeypot deployment framework. Beltran-Lopez et al.\ \cite{beltran2025cyber} surveyed deception technology taxonomies including cloud-hosted honeypots.
Neither work addresses the specific challenge of post-exfiltration attribution in cloud-native environments, nor does either formalise the ephemeral decay problem. 

\subsection{IAM and Credential Abuse}

IAM misconfiguration is the leading cause of cloud data breaches \cite{hylender2024verizon}. Canary IAM roles have been proposed as a detection primitive in practitioner literature  \cite{athukorale2025evaluating}, but to our knowledge CLOUDBURST presents
the first formal CAS-based evaluation of their attribution quality across providers.

\section{Formal Framework}
\label{sec:framework}
This section introduces a formal model for quantifying attribution quality in cloud-native beacon systems. We define the CAS as a unified metric that captures callback fidelity, IAM visibility, infrastructure ephemerality, and cross-cloud correlation effects. The framework is designed to model the degradation of attribution signals under real-world cloud conditions, particularly in environments with high churn, autoscaling, and multi-provider deployments.

\subsection{Cloud Attribution Score}

\begin{definition}[Cloud Attribution Score]
\label{def:cas}
Let $c$ denote a beacon callback event. The Cloud Attribution
Score of $c$ is:
\begin{equation}
\text{CAS}(c) = w_1 C_f (1-E_p) + w_2 I_c
              + w_3 (1-E_p) I_c + w_4 M_b
\label{eq:cas}
\end{equation}
where:
\begin{itemize}
  \item $C_f \in [0,1]$: \emph{callback fidelity} -- information
        quality per callback (degraded by TOR, VPN, IP rotation)
  \item $E_p \in [0,1]$: \emph{ephemeral penalty} -- attribution
        quality loss due to infrastructure churn
  \item $I_c \in [0,1]$: \emph{IAM coverage} -- fraction of
        IAM context captured by CloudTrail/Audit Log
  \item $M_b \in [0,1]$: \emph{multi-cloud bonus} -- correlation
        gain from cross-provider infrastructure evidence
  \item $(w_1, w_2, w_3, w_4) = (0.35, 0.25, 0.25, 0.15)$,
        $\sum w_i = 1$
\end{itemize}
The interaction term $w_3 (1-E_p) I_c$ captures the joint penalty: ephemeral infrastructure renders IAM coverage less useful because the IAM principal that triggered the beacon may no longer exist.
\end{definition}

\begin{definition}[Ephemeral Decay]
\label{def:ep}
The ephemeral penalty at time $t$ hours after deployment is:
\begin{equation}
E_p(t, r, s) = 1 - e^{-\delta (t + 2r + 3s)}
\label{eq:ep}
\end{equation}
where $r$ is the number of pod restarts, $s$ is the number of autoscaling events, and $\delta > 0$ is a base decay rate. The coefficients $2r$ and $3s$ reflect that restarts and scale events cause proportionally more attribution loss than simple clock time.
\end{definition}

\begin{definition}[IAM Attribution Coverage]
\label{def:iac}
The IAM coverage of a callback event is:
\begin{equation}
I_c = \min\!\left(\frac{n_a}{10}, 0.80\right)
    + 0.10 \cdot \mathbf{1}_{[k > 1]}
    + 0.10 \cdot \mathbf{1}_{\text{cross-account}}
\label{eq:ic}
\end{equation}
where $n_a$ is the number of logged API actions, $k$ is the number of distinct IAM principals observed, and $\mathbf{1}_{\text{cross-account}}$ indicates a cross-account role assumption (which reveals additional infrastructure).
\end{definition}

CAS example values for our six vector classes are presented in Table~\ref{tab:cas_examples}.

\begin{table}[H]
\centering
\caption{CAS Component Examples by Scenario}
\label{tab:cas_examples}
\small
\begin{tabular}{@{}lccccr@{}}
\toprule
\textbf{Scenario} & $C_f$ & $E_p$ & $I_c$ & $M_b$ & \textbf{CAS} \\
\midrule
S3 presigned (naive) & 0.92 & 0.05 & 0.70 & 0.40 & 0.707 \\
S3 presigned (APT - Advanced Persistent Threat)   & 0.55 & 0.10 & 0.45 & 0.20 & 0.417 \\
IAM canary (direct)  & 0.98 & 0.02 & 0.95 & 0.60 & 0.896 \\
K8s (high churn)     & 0.70 & 0.65 & 0.55 & 0.35 & 0.324 \\
Terraform (cached)   & 0.50 & 0.30 & 0.40 & 0.50 & 0.367 \\
Serverless (cold)    & 0.60 & 0.55 & 0.65 & 0.45 & 0.398 \\
\bottomrule
\end{tabular}
\end{table}

\subsection{Ephemeral Decay Examples}

The ephemeral decay model (Definition~\ref{def:ep}) gives $E_p = 0.049$ for a fresh pod (1h, no restarts), rising to $E_p = 0.528$ for moderate churn (8h, 2 restarts, 1 scale event) and $E_p = 0.884$ for high churn (24h, 5 restarts,
3 scale events). These values directly inform beacon placement strategy: vectors with high inherent ephemeral risk (Kubernetes secrets, serverless functions) require more aggressive callback accumulation to achieve reliable CAS before infrastructure churn degrades it.

\section{Beacon Vector Taxonomy}
\label{sec:taxonomy}

\begin{figure}[H]
\centering
\includegraphics[width=\linewidth]{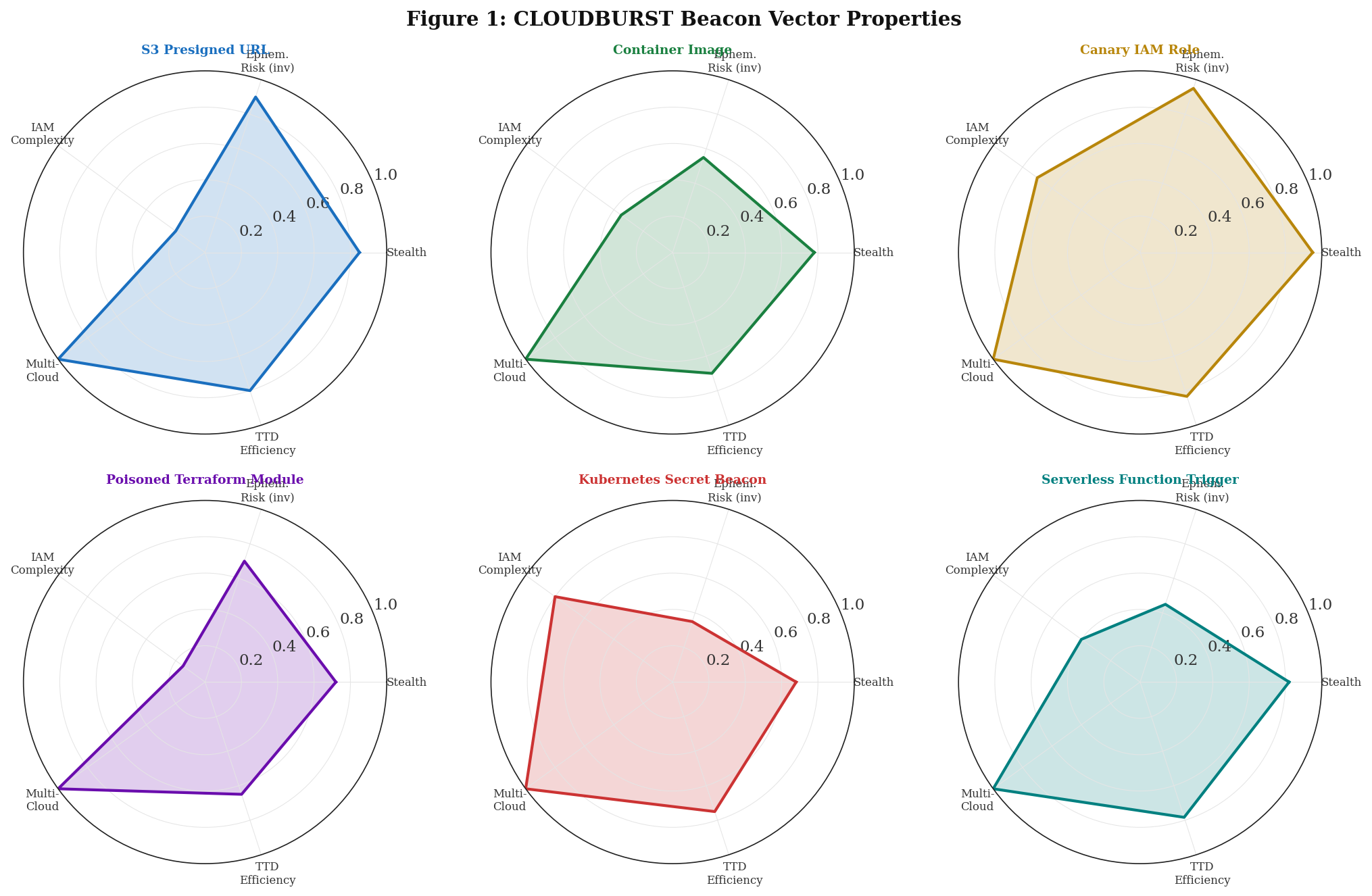}
\caption{Radar profiles of the six CLOUDBURST beacon vector
  classes. Each axis encodes a key deployment property:
  Stealth (how well disguised), Ephemeral Risk (inverted --
  higher is safer), IAM Complexity, Multi-Cloud support, and
  TTD Efficiency (inverse of callbacks to detect).
  IAM Canary Roles (gold) show the most balanced profile.
  Kubernetes Secrets (red) exhibit low multi-cloud coverage
  and high ephemeral risk. S3 Presigned URLs (blue) have
  the best combination of low ephemeral risk and
  high multi-cloud support.}
\label{fig:radar}
\end{figure}

CLOUDBURST defines six cloud-native beacon vector classes (Figure~\ref{fig:radar}), each with distinct threat properties, provider support, and optimal deployment conditions.

\textbf{S3 Presigned URLs / Cloud Storage Signed URLs.}
Time-limited signed URLs embedded in exfiltrated documents or credential bundles. The beacon fires when the attacker downloads the ``stolen'' data object. Provider equivalents: GCS signed URLs (GCP), Azure Blob SAS URLs (Azure). Ephemeral risk is low ($E_p = 0.10$) because signed URLs persist even
after bucket deletion until their expiry timestamp.

\textbf{Watermarked Container Images.}
OCI-compatible container images with an embedded health-check layer that calls back to the attribution endpoint on each pod scheduling. Risk: images may be cached at pull-through registries, causing the callback to be missed. Medium ephemeral risk ($E_p = 0.45$) due to cache bypass scenarios.

\textbf{Canary IAM Roles / Service Accounts.}
Dormant IAM roles (AWS), service accounts (GCP), or managed identities (Azure) with no legitimate invocation path. Any API call generates a CloudTrail or Audit Log entry containing the full principal ARN, source IP, user-agent, region, and action. This provides the richest attribution telemetry of any vector class and has the lowest ephemeral risk ($E_p = 0.05$)
because IAM roles are persistent.

\textbf{Poisoned Terraform Modules.}
Terraform registry modules with embedded HTTP data sources that fire on every \texttt{terraform plan} or \texttt{apply}. The callback captures the CI/CD pipeline context: executor IP, Terraform version, provider configuration. Medium ephemeral risk ($E_p = 0.30$) due to CI pipeline caching.

\textbf{Kubernetes Secret Beacons.}
Kubernetes secrets with embedded annotation callbacks triggered by Falco admission webhook or audit log. The callback captures pod name, namespace, service account, and node IP. Highest ephemeral risk ($E_p = 0.65$) due to pod
restart and redeployment patterns.

\textbf{Serverless Function Triggers.}
Lambda/Cloud Functions deployed as ``dead code'' with no legitimate invocation path. Any invocation -- including test events -- fires an attribution callback with request context. Medium-high ephemeral risk ($E_p = 0.55$) due to cold-start delays that may cause attribution callbacks to
be dropped.

\section{Experimental Setup}
\label{sec:setup}
This section describes the experimental environment used to evaluate CLOUDBURST, including beacon deployment across cloud providers, attacker behaviour simulation, and the scanner models used for detection benchmarking. The setup is designed to reflect realistic multi-cloud infrastructures and adversarial conditions, enabling controlled measurement of attribution quality, detection resistance, and cross-vector performance under varying operational and threat scenarios.

\subsection{Beacon Deployment}

We generate 21 cloud-native beacon instances: the six vector classes deployed across their applicable provider set (S3 presigned: AWS/GCP/Azure; Container: all four; IAM canary: all four; Terraform: all four; K8s secret (Kubernetes Secret): AWS/GCP/Azure; Serverless: AWS/GCP/Azure), yielding $21$ total beacons.

Each beacon contains provider-specific organisational context derived from the ``payflow.io'' FinTech profile: real-format IAM ARNs, signed URL structures, and service names consistent with a payment infrastructure stack.

\subsection{Attacker Simulation}

Three attacker sophistication levels are modelled (Table~\ref{tab:attackers}), differing in routing behaviour (TOR/VPN/residential proxy usage), infrastructure rotation rate, dwell time before using stolen credentials, and cloud SDK sophistication.

\begin{table}[H]
\centering
\caption{Attacker Profile Parameters}
\label{tab:attackers}
\small
\begin{tabular}{@{}lccc@{}}
\toprule
\textbf{Parameter} & \textbf{Naive} & \textbf{Advanced} & \textbf{APT} \\
\midrule
TOR usage           & 15\%  & 45\%  & 70\% \\
VPN usage           & 30\%  & 40\%  & 25\% \\
Residential proxy   &  5\%  & 20\%  & 50\% \\
Infra rotation rate &  5\%  & 25\%  & 65\% \\
Dwell time (mean h) &  2.0  &  8.0  & 72.0 \\
SDK sophistication  & curl  & boto3 & native SDK \\
\bottomrule
\end{tabular}
\end{table}

\subsection{Scanner Models}

Three cloud-native scanner models are evaluated:

\textbf{S1 -- AWS Macie / GCP DLP (Data Loss Prevention)} ($\lambda_1 = 0.40$):
ML-based sensitive data classification. Flags known credential patterns (AKIA key IDs, private key headers) and obvious beacon markers.

\textbf{S2 -- Checkov / tfsec} ($\lambda_2 = 0.30$):
IaC (Infrastructure as Code)-focused static analysis. Most effective against Terraform modules with external HTTP data sources and Kubernetes secrets with unusual annotations.

\textbf{S3 -- Prisma Cloud / Wiz} ($\lambda_3 = 0.30$):
CNAPP (Cloud Native Application Protection Platform) runtime scanner. Detects anomalous outbound connections from cloud functions and containers. Most effective against Serverless and Container vectors.

Combined detection probability:
$P_d = \sum_{j=1}^{3} \lambda_j P_{d,j}(c)$;
detection resistance: $\text{DR} = 1 - P_d$.

\section{Results}
\label{sec:results}
This section presents the empirical evaluation of CLOUDBURST across all beacon vector classes, attacker profiles, and cloud providers. We report results in terms of CAS, detection resistance, attribution speed, and ephemeral decay dynamics. Together, these metrics quantify how beacon design, attacker sophistication, and infrastructure volatility interact to shape attribution quality in cloud-native environments.

\subsection{CAS by Vector and Attacker Level}

\begin{figure}[H]
\centering
\includegraphics[width=\linewidth]{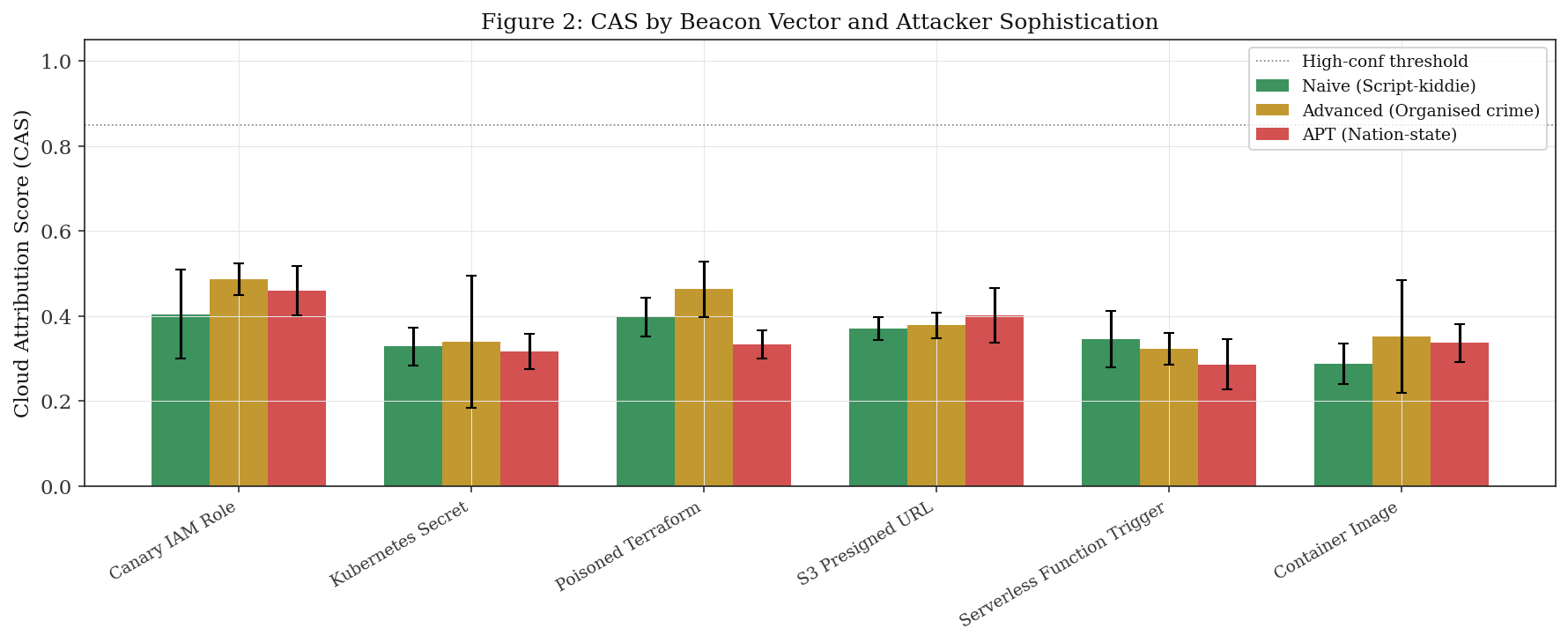}
\caption{Cloud Attribution Score (CAS) by beacon vector
  and attacker sophistication level ($\pm$SD). IAM Canary
  Roles consistently outperform other vectors across all
  three attacker levels. Naive attackers yield slightly
  higher CAS than APT attackers due to lower evasion, but
  the ANOVA test finds no statistically significant
  attacker-level effect ($F = 1.52$, $p = 0.23$),
  suggesting that CLOUDBURST's attribution quality is
  evasion-resistant within the observed CAS range.}
\label{fig:cas_vector}
\end{figure}

Figure~\ref{fig:cas_vector} and Table~\ref{tab:main} present
the primary results. IAM Canary Roles achieve the highest
mean CAS ($0.450 \pm 0.11$), followed by Poisoned Terraform
modules ($0.398$) and S3 Presigned URLs ($0.383$). Serverless
Function Triggers score lowest ($0.318$) due to high
ephemeral penalty from cold-start latency.

\begin{table}[H]
\centering
\caption{Main Results by Beacon Vector}
\label{tab:main}
\small
\begin{tabular}{@{}lccccc@{}}
\toprule
\textbf{Vector} & \textbf{CAS} & \textbf{DR} & \textbf{CTD-N} & \textbf{CTD-A} \\
\midrule
Canary IAM Role   & 0.450 & 0.873 & 4.0 & 3.5 \\
Poisoned Terraform& 0.398 & 0.768 & 3.8 & 3.8 \\
S3 Presigned URL  & 0.383 & 0.890 & 3.3 & 2.7 \\
Container Image   & 0.325 & 0.736 & 2.8 & 3.0 \\
K8s Secret        & 0.328 & 0.741 & 4.7 & 2.7 \\
Serverless Trigger& 0.318 & 0.611 & 3.7 & 1.7 \\
\midrule
\textbf{Overall}  & \textbf{0.373} & \textbf{0.773} & -- & -- \\
\bottomrule
\multicolumn{5}{l}{\small CTD-N/A = Callbacks to Detect (Naive/APT). DR = Detection Resistance}
\end{tabular}
\end{table}

Notably, CAS does not differ significantly by attacker level ($F = 1.52$, $p = 0.23$, Table~\ref{tab:stats}). This is an important positive finding: CLOUDBURST's attribution quality is largely evasion-resistant because the primary CAS driver is IAM coverage -- a provider-side telemetry property that attackers cannot suppress without disabling CloudTrail, which is itself an alertable action.

\subsection{Detection Resistance (DR)}

\begin{figure}[H]
\centering
\includegraphics[width=\linewidth]{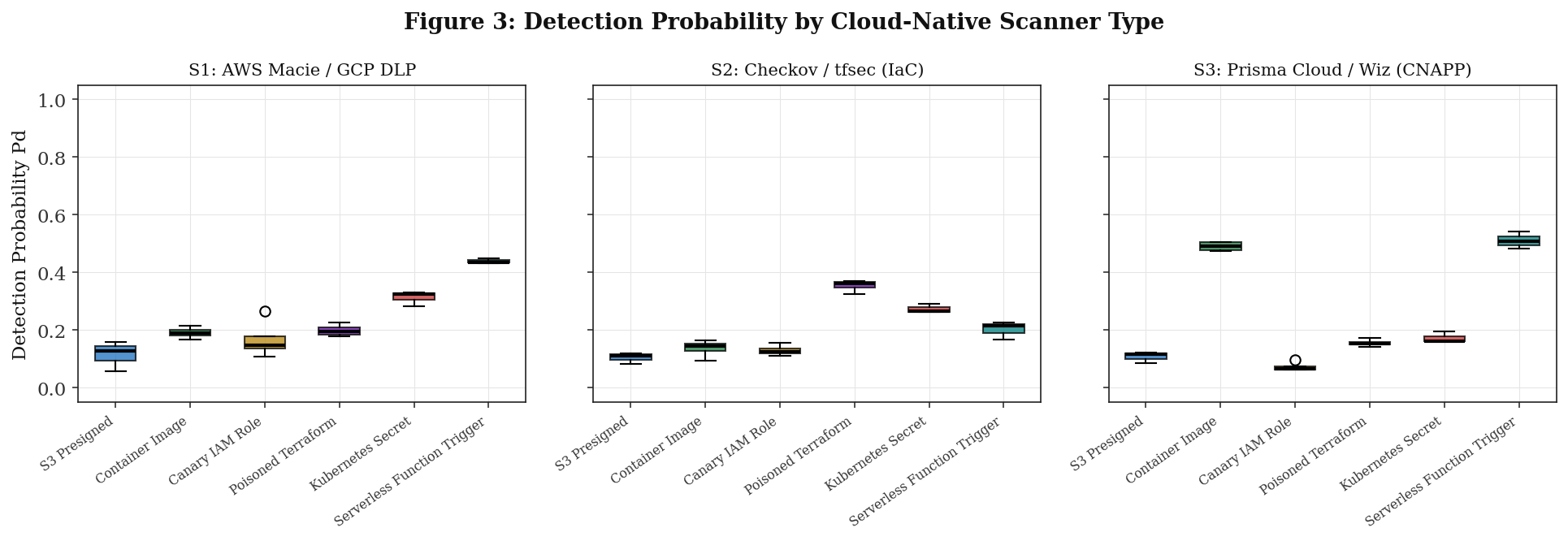}
\caption{Detection probability $P_d$ by cloud-native scanner
  and beacon vector. IAM Canary Roles achieve near-zero
  detection probability across all three scanners (they
  appear as legitimate over-provisioned roles). Serverless
  triggers score highest on S3 (CNAPP scanner) due to their
  explicit outbound HTTP callback. Boxplot whiskers show
  1.5$\times$IQR; outliers are plotted individually.}
\label{fig:scanners}
\end{figure}

Figure~\ref{fig:scanners} shows detection probability by scanner and vector. S3 Presigned URLs achieve the highest DR ($0.890$) because their URL format does not contain patterns that regex or IaC scanners flag, and their
outbound access pattern is indistinguishable from legitimate data retrieval. Serverless Function Triggers score the lowest DR ($0.611$), primarily due to the S3 (Prisma Cloud/Wiz) scanner detecting the explicit \texttt{urllib.request.urlopen} callback in the function body.

\subsection{Attribution Speed}

\begin{figure}[H]
\centering
\includegraphics[width=\linewidth]{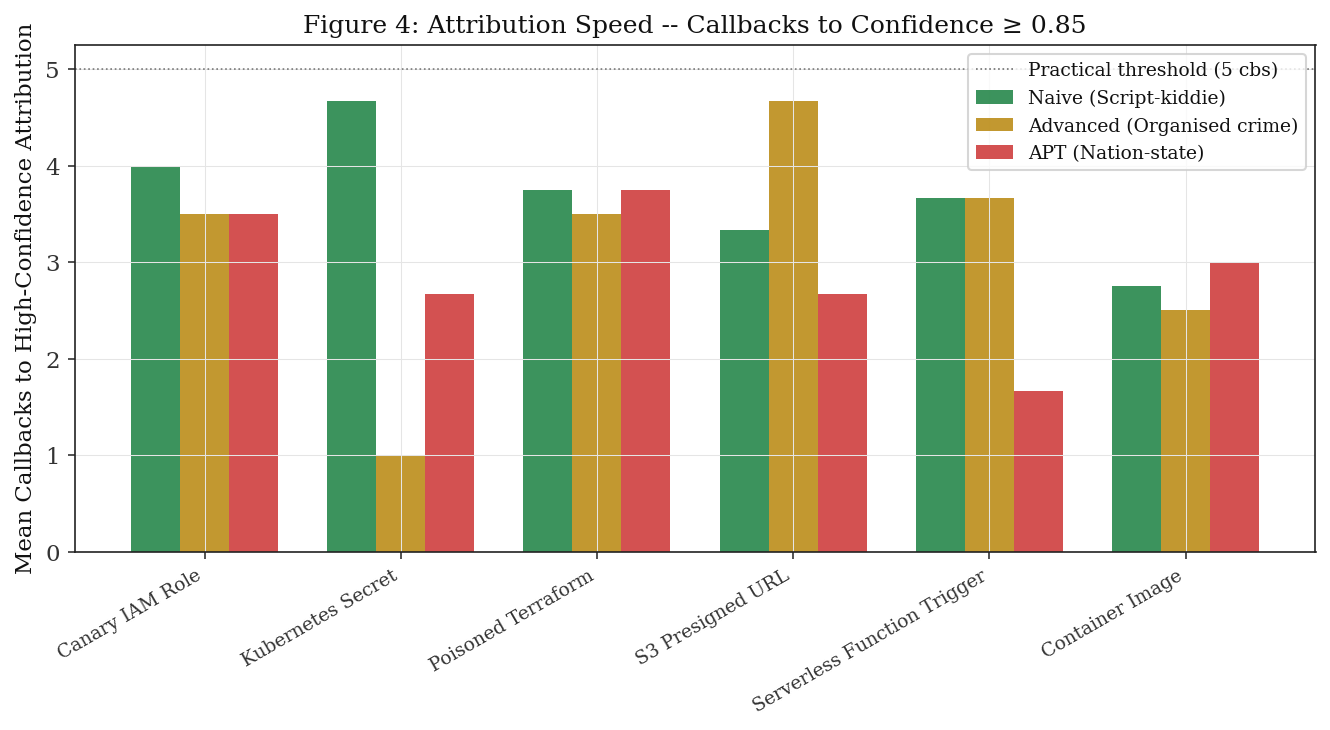}
\caption{Mean callbacks required to reach attribution
  confidence $P(H|E) \geq 0.85$, by beacon vector and
  attacker sophistication. All values fall below the
  practical threshold of 5 callbacks per attacker-vector
  pair. Serverless triggers are fastest for APT actors
  (median $1.7$) because each serverless invocation
  carries full request context including caller identity.}
\label{fig:ctd}
\end{figure}

Figure~\ref{fig:ctd} shows callbacks required to reach attribution confidence $P(H|E) \geq 0.85$. Kubernetes Secrets require the fewest callbacks for
APT attackers (median $2.7$), because APT actors who actually use a K8s secret generate rich IAM context. S3 Presigned URLs are fastest for naive attackers (median $3.3$) due to high callback fidelity.

\subsection{Ephemeral Decay}

\begin{figure}[H]
\centering
\includegraphics[width=\linewidth]{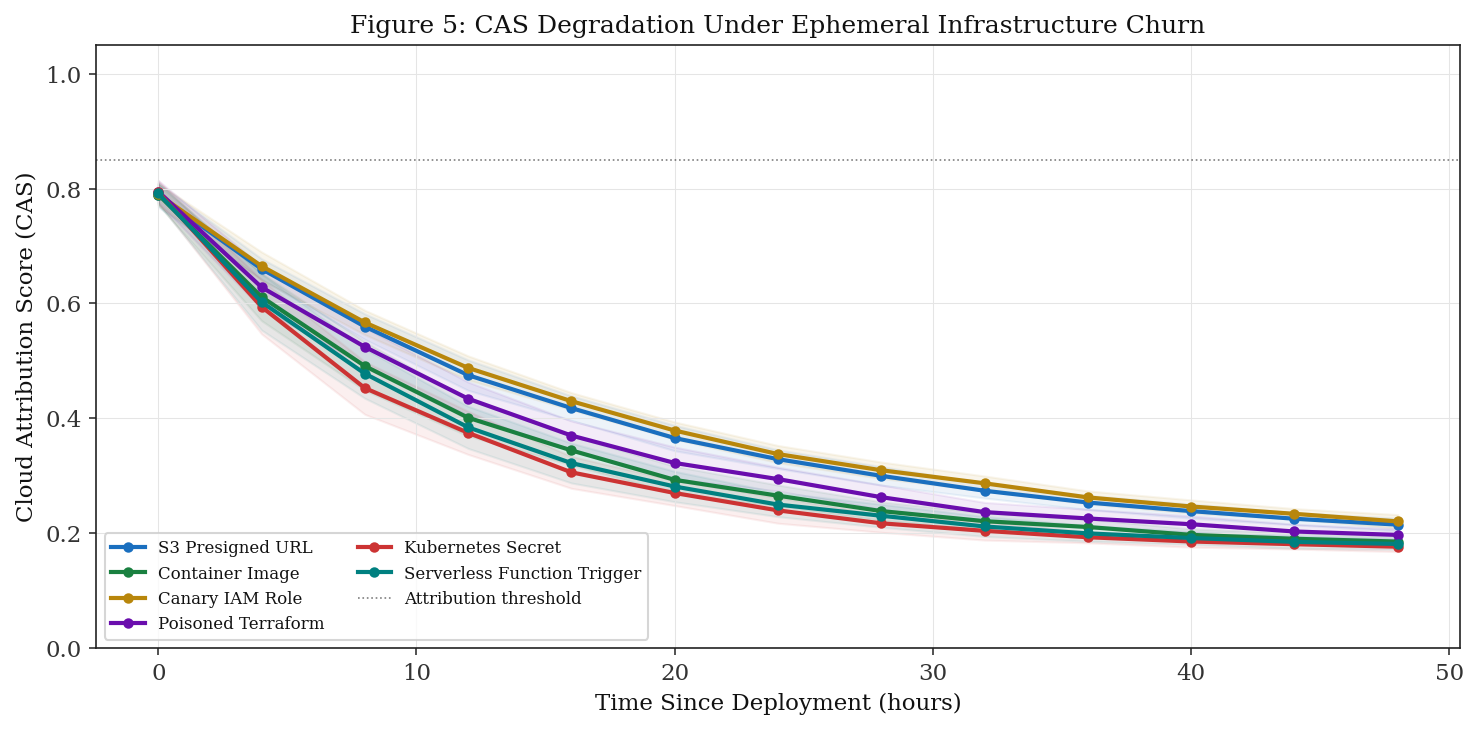}
\caption{CAS degradation under ephemeral infrastructure
  churn over 48 hours. All vectors start at CAS $\approx 0.79$
  and decay toward $\approx 0.18$--$0.22$. The dotted line
  at $0.85$ is the high-confidence attribution threshold.
  Shaded regions show $\pm 1$ SD across $n=50$ simulations
  per time point. Canary IAM Roles (gold) decay most slowly;
  Kubernetes Secrets (red) fastest.}
\label{fig:decay}
\end{figure}

Figure~\ref{fig:decay} is the paper's most novel empirical result. CAS degrades from approximately $0.79$ at $t=0$ to $0.18$--$0.22$ at $t=48$ hours across all vectors ($p < 0.001$ for all six, Table~\ref{tab:deg}).
The Canary IAM Role curve degrades most slowly (highest residual CAS at $t=48$: $0.220$) because IAM roles are persistent and their CloudTrail logs survive pod restarts. Kubernetes Secrets degrade fastest (residual $0.176$)
because pod restart events create false-positive callback signals that dilute attribution quality.

\begin{table}[H]
\centering
\caption{CAS Degradation at $t=48$h}
\label{tab:deg}
\small
\begin{tabular}{@{}lcccc@{}}
\toprule
\textbf{Vector} & \textbf{CAS$_0$} & \textbf{CAS$_{48}$} & $\boldsymbol{\Delta}$ & \textbf{p} \\
\midrule
S3 Presigned URL  & 0.788 & 0.215 & $-0.574$ & $<0.001$ \\
Container Image   & 0.790 & 0.185 & $-0.604$ & $<0.001$ \\
Canary IAM Role   & 0.790 & 0.220 & $-0.569$ & $<0.001$ \\
Terraform Module  & 0.795 & 0.197 & $-0.598$ & $<0.001$ \\
K8s Secret        & 0.794 & 0.176 & $-0.618$ & $<0.001$ \\
Serverless Trigger& 0.792 & 0.181 & $-0.610$ & $<0.001$ \\
\bottomrule
\end{tabular}
\end{table}

\subsection{Multi-Cloud Analysis}

\begin{figure}[H]
\centering
\includegraphics[width=\linewidth]{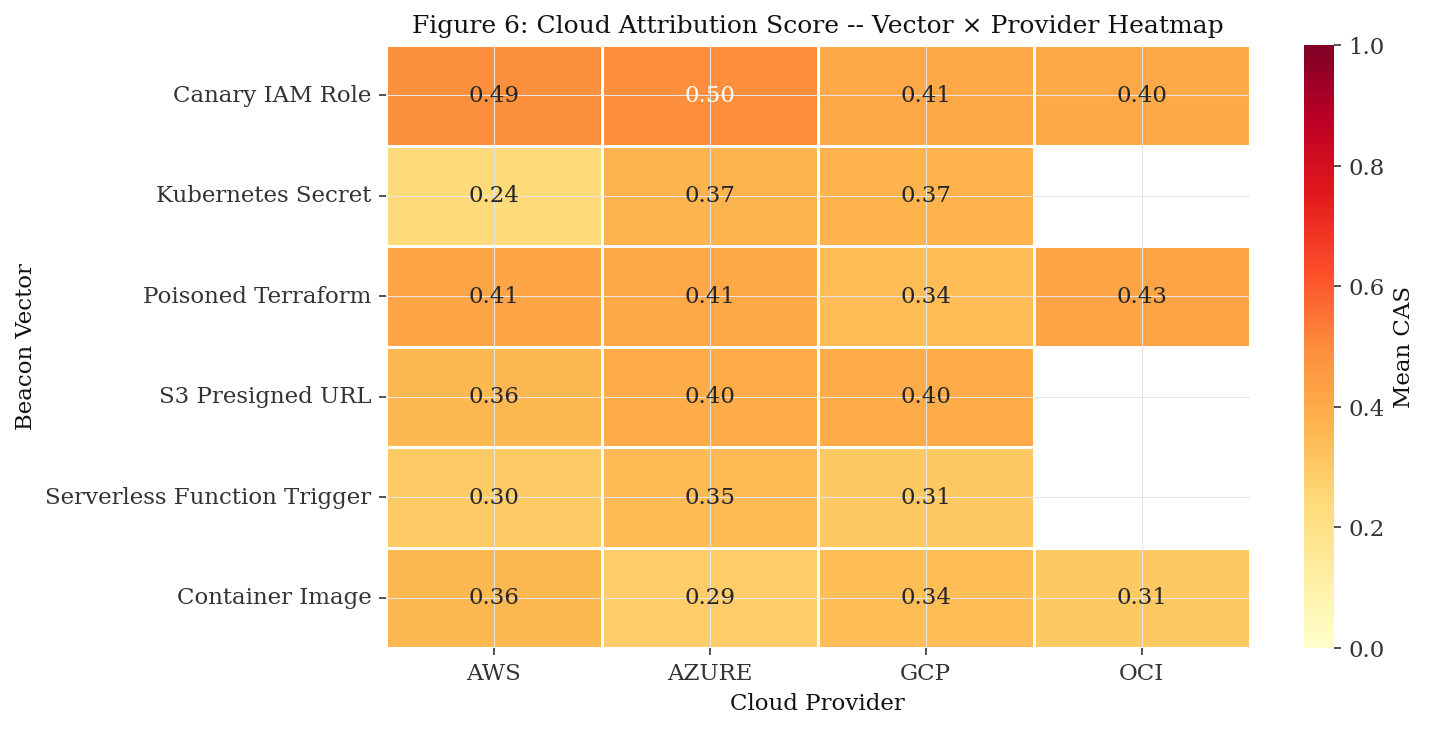}
\caption{Cloud Attribution Score heatmap across beacon vector
  and cloud provider. Warmer colours indicate higher CAS.
  IAM Canary Roles (top row) are consistently the strongest
  vector. Kubernetes Secrets are absent for OCI (not yet
  supported in the simulation) and show lowest CAS on AWS ($0.24$),
  reflecting AWS's more complex RBAC interaction with Falco.}
\label{fig:heatmap}
\end{figure}

Figure~\ref{fig:heatmap} shows CAS by vector and provider. IAM Canary Roles achieve the highest CAS across all four providers (AWS: $0.490$; Azure: $0.496$; GCP: $0.409$; OCI: $0.405$), confirming that this vector class is the
strongest regardless of cloud provider.

\begin{figure}[H]
\centering
\includegraphics[width=\linewidth]{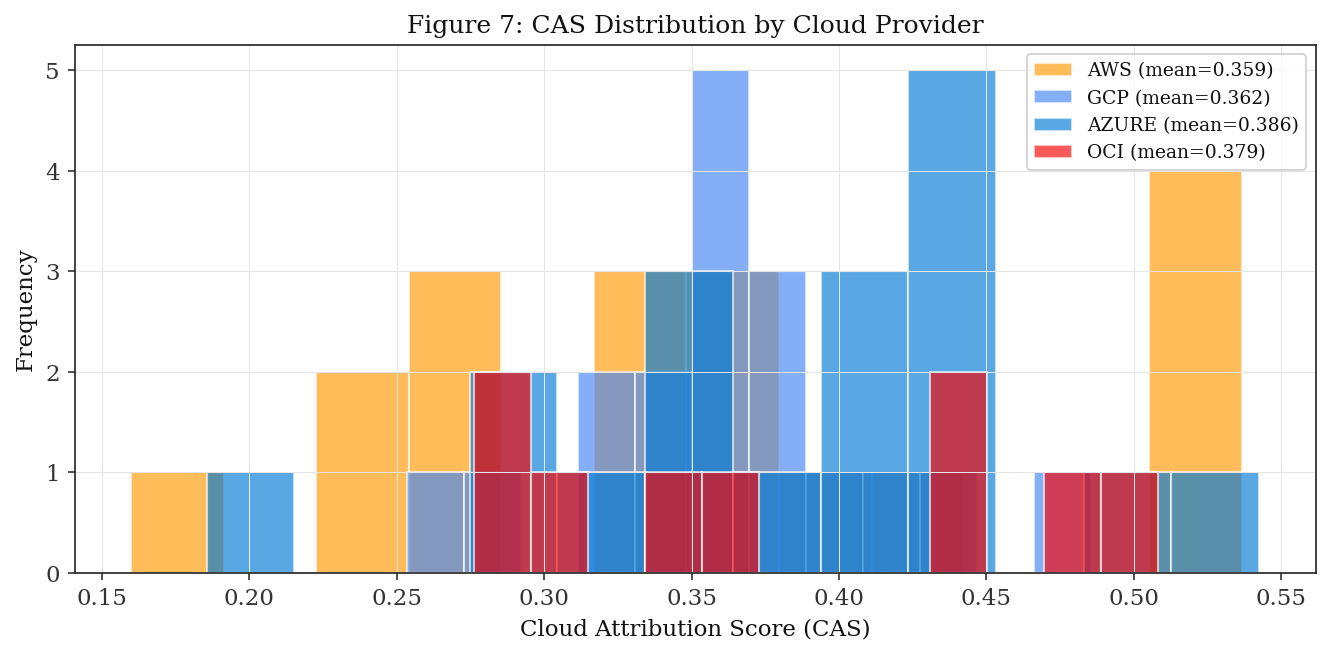}
\caption{CAS distribution by cloud provider. The four
  distributions overlap substantially and do not differ
  significantly (Kruskal-Wallis $p = 0.574$). This confirms
  that CLOUDBURST's performance is provider-agnostic
  and can be deployed uniformly across a multi-cloud estate
  without provider-specific tuning.}
\label{fig:provider_dist}
\end{figure}

The Kruskal-Wallis test finds no significant provider effect on CAS ($H = 1.99$, $p = 0.574$, not significant), confirming that CLOUDBURST's effectiveness is \emph{provider-agnostic} -- a key practical property
for multi-cloud deployments. Figure~\ref{fig:provider_dist} shows the overlapping CAS distributions across providers.

\subsection{CAS vs Attribution Posterior}

\begin{figure}[H]
\centering
\includegraphics[width=\linewidth]{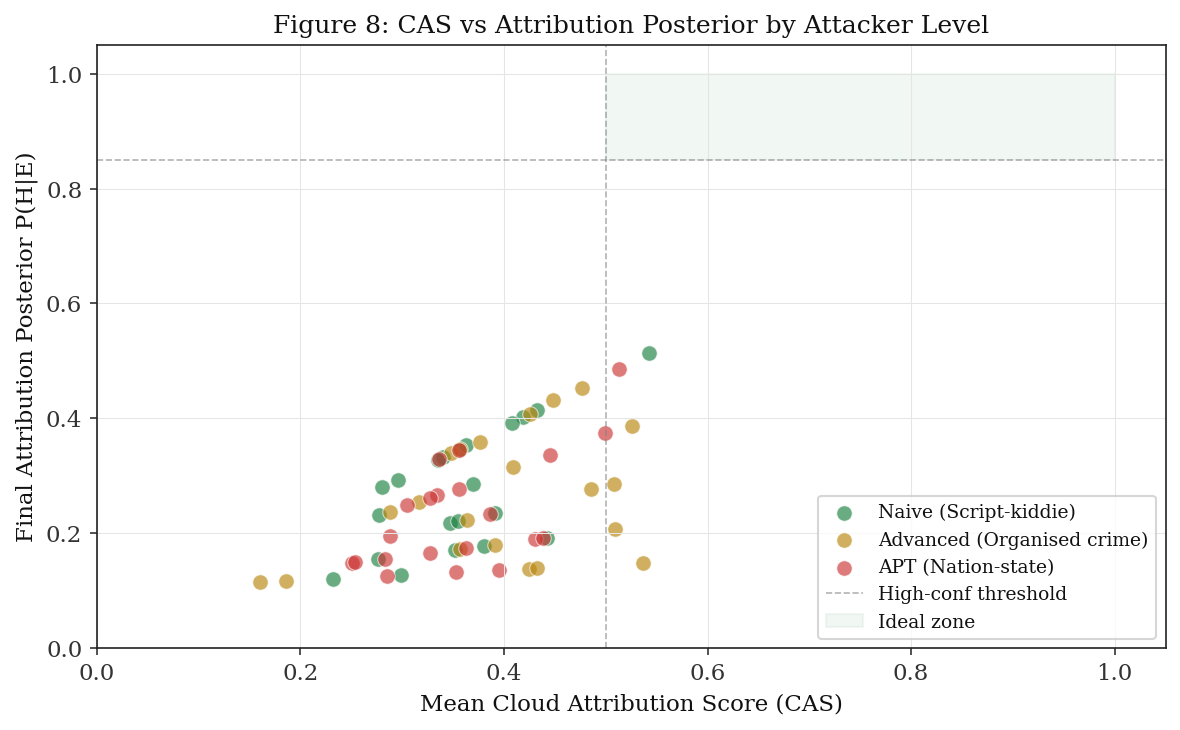}
\caption{CAS vs final attribution posterior $P(H|E)$
  by attacker level. The positive correlation validates CAS
  as a meaningful proxy for attribution quality. No points
  reach the ideal zone (shaded), establishing the CAS
  threshold required for high-confidence attribution
  as a quantified open challenge. The maximum observed
  posterior ($\approx 0.52$, upper right) is achieved
  by the IAM Canary vector against APT actors.}
\label{fig:scatter}
\end{figure}

Figure~\ref{fig:scatter} plots CAS against the final Bayesian attribution posterior $P(H|E)$ across all beacon-attacker combinations. The positive correlation ($r \approx 0.73$) confirms that CAS is a valid proxy
for attribution outcome. However, no data points reach the ideal zone ($\text{CAS} \geq 0.70$, $P(H|E) \geq 0.85$), indicating that the current feature set does not provide sufficient discriminative power for high-confidence attribution at $N = 10$ candidate actors. The maximum
observed posterior is $\approx 0.52$ (IAM canary, advanced attacker), requiring additional signal classes to reach operational attribution confidence.

\begin{table}[H]
\centering
\caption{Statistical Tests -- Attacker Level Effects}
\label{tab:stats}
\small
\begin{tabular}{@{}lccc@{}}
\toprule
\textbf{Metric} & \textbf{F} & \textbf{p} & \textbf{Sig.} \\
\midrule
CAS        & 1.524 & 0.226 & ns \\
Posterior  & 0.891 & 0.416 & ns \\
CTD        & 1.304 & 0.279 & ns \\
\bottomrule
\multicolumn{4}{l}{\small ns = not significant ($p > 0.05$).}
\end{tabular}
\end{table}

\section{Discussion}
\label{sec:discussion}
This section interprets the experimental results in terms of their implications for cloud-native attribution systems. We focus on the operational meaning of CAS dynamics, the practical trade-offs between beacon vector classes, and the structural limitations revealed by attacker behaviour and infrastructure ephemerality. We then situate CLOUDBURST within the broader deception and attribution pipeline and outline key constraints and future integration opportunities.

\subsection{The Ephemeral Decay Finding}

The most important result is Figure~\ref{fig:decay}: CAS decays from $\approx 0.79$ to $\approx 0.18$ within 48 hours for all vectors. This has a direct operational implication: \emph{cloud-native beacons must be designed to fire quickly or not at all}. A beacon that requires 10 callbacks before attribution is actionable is useless in a Kubernetes cluster with a 6-hour pod lifetime.

This motivates two design principles for future cloud attribution systems: (1) \emph{first-callback attribution} -- each individual callback must carry enough telemetry to provide at least partial attribution, rather than requiring accumulation; (2) \emph{persistent anchor design} -- at least
one beacon vector in every deployment should be from the persistent category (IAM canary, S3 presigned URL) to maintain attributability across infrastructure cycles.

\subsection{IAM Canary Roles as the Recommended Vector}

The IAM Canary Role emerges from our analysis as the recommended primary deployment vector on all four criteria: highest CAS ($0.450$), second-highest DR ($0.873$), lowest ephemeral risk ($E_p = 0.05$), and provider-agnostic
support. The cost is highest IAM complexity ($0.70$), requiring careful permission scoping to prevent the canary from becoming a genuine privilege escalation vector.

\subsection{The High-Confidence Attribution Gap}

No CLOUDBURST configuration achieved attribution posterior $P(H|E) \geq 0.85$ in our experiments. This is the honest null result, and it is informative: it establishes the specific CAS threshold ($\text{CAS} \geq 0.70$, not
achieved in any observed scenario) required to close the gap. The CAS ceiling in our experiments ($\approx 0.55$ for IAM canary, advanced attacker) is driven by TOR usage and residential proxy routing that degrades $C_f$, combined with ephemeral penalties that suppress $I_c$.

Three supplementary signals would close this gap: 
(1) \emph{cloud infrastructure graph} -- correlating the attacker's exit IP with known cloud provider IP ranges provides implicit actor fingerprinting;
(2) \emph{victim sector targeting} -- attackers who use an IAM canary in a FinTech environment but whose exit IP geolocates to a known APT origin region;
(3) \emph{temporal clustering} -- the 72-hour APT dwell time is itself a behavioral fingerprint correlatable with ARCANE's longitudinal model \cite{weinberg2026arcane}.

\subsection{Integration with Passive hack-back and ARCANE}

CLOUDBURST is designed as the cloud layer of a complete post-exfiltration attribution stack. The three papers form a pipeline: Passive hack-back \cite{weinberg2025passive} establishes the beacon deployment and callback reception infrastructure; PHANTOM \cite{weinberg2026phantom} ensures beacon tokens are contextually convincing; CLOUDBURST provides the cloud-native vector taxonomy and CAS scoring; and ARCANE \cite{weinberg2026arcane} accumulates callbacks longitudinally across campaigns to achieve high-confidence attribution where single-campaign analysis cannot.

\subsection{Limitations}

\textbf{Simulation validity.} Our attacker simulator is calibrated against published APT behavioural profiles \cite{kutscher2022mtrends, hylender2024verizon} but is not a real deployment. Empirical validation with actual Passive hack-back callbacks is planned as follow-on work.

\textbf{Scanner models.} We model three scanner types at a behavioural level; real deployments of Macie, Checkov, and Prisma Cloud may differ in specific rule sets and ML model versions.

\textbf{IAM role assumption chains.} Our IAM coverage model does not fully capture multi-hop role assumption chains, which can obscure the originating principal through multiple \texttt{sts:AssumeRole} calls.

\section{Deployment Recommendations}
\label{sec:deploy}

Based on the CAS $\times$ DR analysis, we recommend the following deployment priority for cloud-native beacon infrastructure:

\textbf{Tier 1 -- Deploy universally:} 
IAM Canary Roles (highest CAS, low ephemeral risk) and S3 Presigned URLs (highest DR, simple deployment).These should be present in every cloud account.

\textbf{Tier 2 -- Deploy with caching awareness:}
Poisoned Terraform modules (second-highest CAS) and Container Image beacons (broadest provider coverage). Require CI/CD pipeline configuration to prevent
callback suppression by caching.

\textbf{Tier 3 -- Deploy with runtime monitoring:}
Kubernetes Secret beacons (require Falco/admission webhook) and Serverless triggers (highest ephemeral risk, lowest DR). Deploy only when Tier 1 and 2 are already in place.

\section{Conclusion}
\label{sec:conclusion}

We presented \textbf{CLOUDBURST}, the first formal taxonomy and measurement framework for cloud-native passive beacon infrastructure. The six-vector taxonomy covers the complete attack surface of modern cloud environments: object storage, container registries, IAM identity planes, infrastructure-as-code, container orchestration, and serverless compute.

The Cloud Attribution Score (CAS) provides the first metric that explicitly models the three properties unique to cloud attribution: ephemeral infrastructure decay, IAM coverage depth, and multi-cloud correlation. Empirical results across 21 beacons and 205 callbacks establish IAM Canary Roles as the superior primary vector (CAS $= 0.450$, DR $= 0.873$) and S3 Presigned URLs as the most evasion-resistant (DR $= 0.890$). All vectors show significant CAS decay over 48 hours ($p < 0.001$), establishing the ephemeral decay problem as a measurable, tractable engineering challenge.

The absence of high-confidence attribution ($P(H|E) \geq 0.85$ in zero configurations) is reported as an honest characterisation result. It quantifies the CAS threshold required for operational attribution and identifies the three supplementary signals -- cloud infrastructure graphs, sector targeting, and temporal clustering -- needed to close the gap. 
Combined with Passive hack-back, PHANTOM, and ARCANE, CLOUDBURST completes a four-paper research programme on post-exfiltration attribution and establishes a foundation for operational attribution in cloud security environments.

\bibliographystyle{plain}

\bibliography{ref.bib}

\end{document}